\documentclass[a4paper,11pt]{scrartcl}
\usepackage{ILD}
\usepackage{tagging}
\usetag{ILD}
\usepackage{subcaption}
\usepackage{hyperref}
\usepackage{wasysym}
\usepackage[backend=biber,style=numeric-comp,sorting=none,mcite=true,doi=false,subentry]{biblatex}
\usepackage{slashed}
\usepackage{wrapfig,rotating}
\usepackage{pgfpages}
\usepackage{fancybox,graphicx}
\usepackage{graphbox}
\usepackage{epstopdf}
\AppendGraphicsExtensions{.eps.gz}
\title{
  New physics searches with the ILD detector at the ILC

}
\ildproc{phys}{2022}{008}
\localrep{DESY-22-063\\}
\date{\today}

\addauthor{
  Mikael Berggren

}
{\institute{1}}
\addinstitute{1}{
    Deutsches Elektronen-Synchrotron DESY,\\
  Notkestr. 85, 22607 Hamburg, Germany

}
\onbehalfof{
    \tagged{POS}{On behalf of }
the ILD concept group.
\tagged{POS}{The support of the LCC generator group, the ILD
software working group, the EGI federation and the Open Science GRID (via
the ILC virtual organisation) is acknowledged.}

}

\abstract{
  Although the LHC experiments have searched for and excluded many proposed new
particles up to masses close to 1 TeV, there are many scenarios that are
difficult to address at a hadron collider. This talk will review a number of
these scenarios and present the expectations for searches at an electron-positron
collider such as the International Linear Collider. The cases discussed include
the light Higgsino, the stau lepton in the coannihilation region relevant to
dark matter, and heavy vector bosons coupling to the s-channel in e$^+$e$^-$ annihilation.
The studies are based on the ILD concept at the ILC.

}

\titlecomment{%
  Presented at the 30th International Symposium on Lepton Photon Interactions at High Energies, hosted by the University of Manchester, 10-14 January 2022.
}

\def\leqsim{\mathbin{\;\raise1pt\hbox{$<$}\kern-8pt\lower3pt\hbox{$\sim$}\;}}
\def\geqsim{\mathbin{\;\raise1pt\hbox{$>$}\kern-8pt\lower3pt\hbox{$\sim$}\;}}
\def\XN#1{\mbox{$ \tilde{\chi}^0_#1                                     $}}

\def\p#1{\mbox{$ \mbox{\bf p}_1                                         $}}

\newcommand{\smur}    {\mbox{$ \tilde{\mu}_{\mathrm R}                     $}}

\newcommand{\stau}    {\mbox{$ \tilde{\tau}                                $}}
\newcommand{\stone}   {\mbox{$ \tilde{\tau}_1                              $}}

\newcommand{\eeto}    {\mbox{$ {\, \mathrm e}^+ {\mathrm e}^- \to             $}}

\newcommand{\ba}{\begin{array}}
\newcommand{\ea}{\end{array}}
\newcommand{\bc}{\begin{center}}
\newcommand{\ec}{\end{center}}
\newcommand{\be}{\begin{eqnarray}}
\newcommand{\eeq}{\end{eqnarray}}
\newcommand{\bes}{\begin{eqnarray*}}
\newcommand{\ees}{\end{eqnarray*}}
\newcommand{\Kz}{\ifmmode {\rm K^0_s} \else ${\rm K^0_s} $ \fi}
\newcommand{\Zz}{\ifmmode {\rm Z^0} \else ${\rm Z^0 } $ \fi}
\newcommand{\xxbar}{\ifmmode {\rm x\bar{x}} \else ${\rm x\bar{x}} $ \fi}
\newcommand{\rphi}{\ifmmode {\rm R\phi} \else ${\rm R\phi} $ \fi}
\def    \missEt      {\ifmmode{/\mkern-11mu E_t}\else{${/\mkern-11mu E_t}$}\fi}
\def    \missE       {\ifmmode{/\mkern-11mu E}\else{${/\mkern-11mu E}$}\fi}
\def    \missp       {\ifmmode{/\mkern-11mu p}\else{${/\mkern-11mu p}$}\fi}
\def    \misspt      {\ifmmode{/\mkern-11mu p_t}\else{${/\mkern-11mu p_t}$}\fi}

\begin{document}
\titlepage

\section{ILC and BSM}
The ILC - the International Linear Collider - is a power-efficient $e^+e^-$ collider with initial $E_{CMS}$ = 250 GeV, 
upgradable up to 1 TeV \cite{Bambade:2019fyw,Behnke:2013xla}. 
As electrons are point-like objects, the initial
    state   is known.
%$\Rightarrow$ missing energy known: essential for SUSY.
    Since the production mechanism is electro-weak, the backgrounds will be low
    meaning that the detectors can be thin and have close to $\sim 4\pi$ coverage, and
    can be operated without trigger.
   In addition, at ILC, both beams will be polarised.
   This combination of polarised beams, low background, known in-state, hermetic detectors and energy upgradability makes
   the ILC the  ideal
  environment for Beyond the Standard Model (BSM) searches \cite{Fujii:2017ekh}.
 %\item Construction under political consideration in Japan.

\subsection{BSM at ILC: the SUSY case}
SUSY is the most complete theory of BSM.
It also serves as a boiler-plate for BSM, since almost any new topology can be obtained in some flavour of SUSY.
In addition it is the most studied model with serious simulation: In most cases presented here, used full simulation of 
the International Large Detector concept at ILC (the ILD \cite{ILDConceptGroup:2020sfq}), with all SM backgrounds, and all
      beam-induced backgrounds included.

\begin{figure}[b]
%\begin{center}
\subfloat[][]{
\includegraphics[scale=0.28]{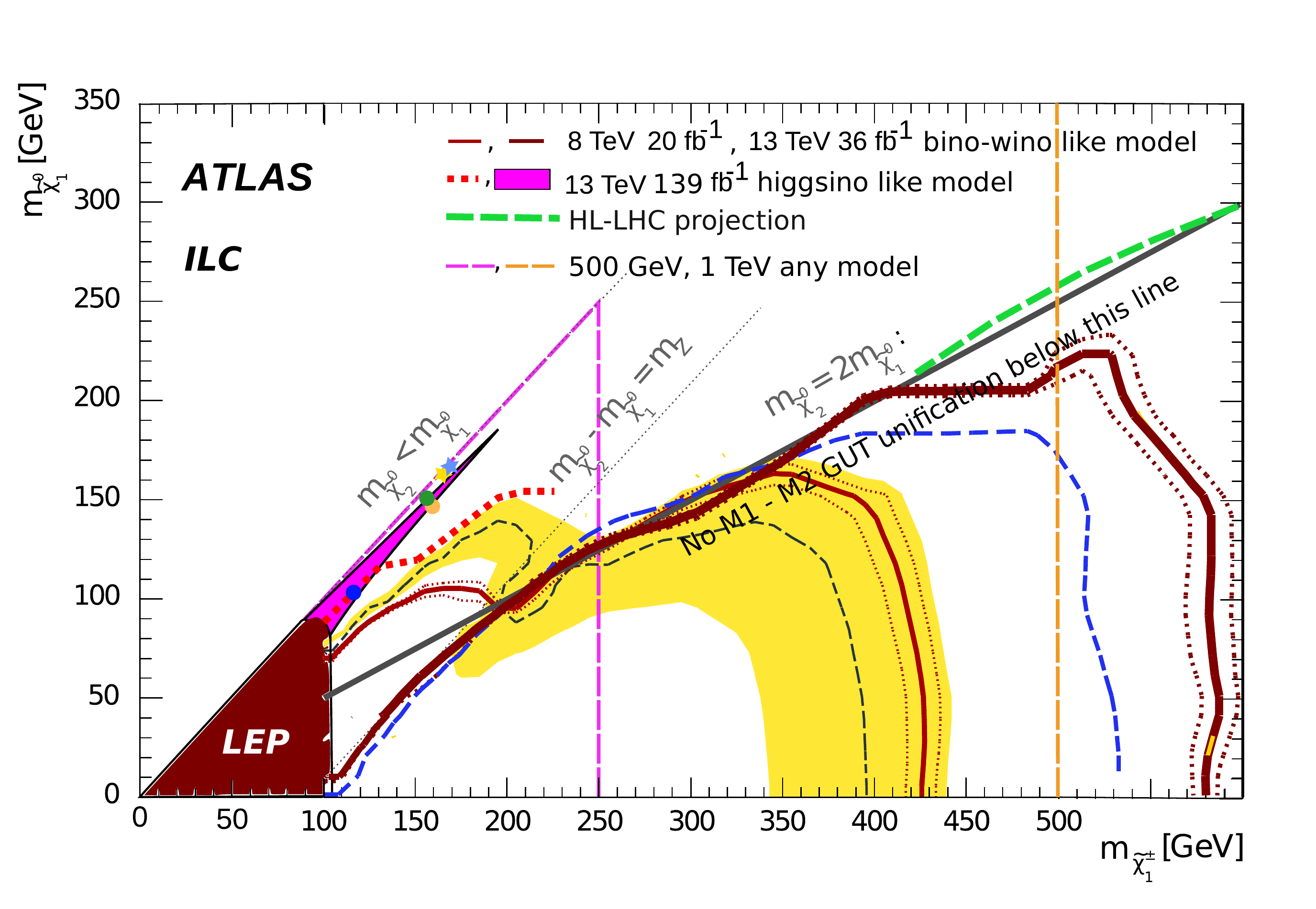}
}
\subfloat[][]{
\includegraphics[scale=0.28]{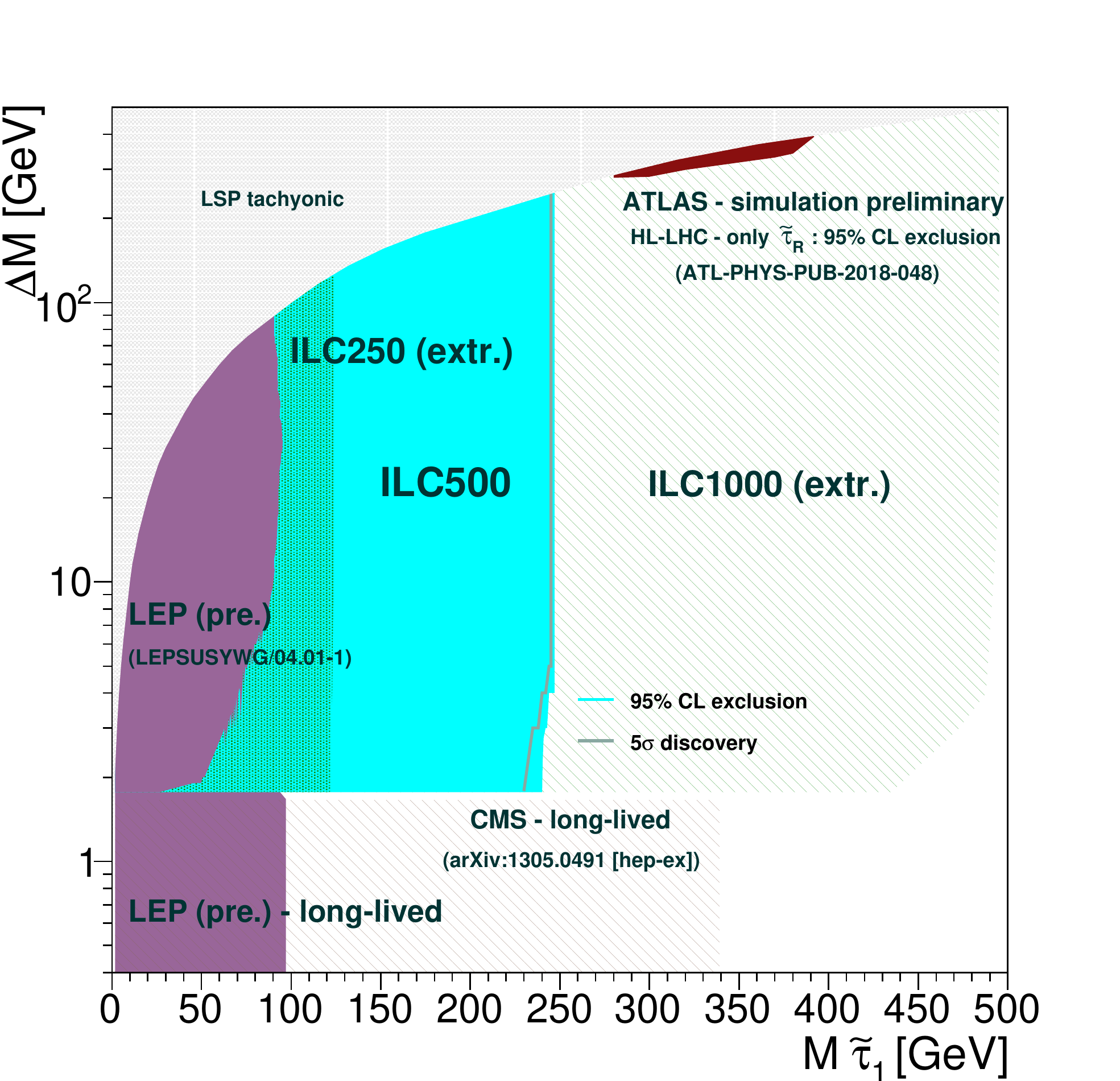}
}
\caption{The expected exclusion and \textit{discovery} reaches at ILC for bosinos (a)
  % \cite{Berggren:2021sns,PardodeVera:2020zlr}
  \cite{ILCInternationalDevelopmentTeam:2022izu}
  and $\stau$:s (b)\cite{deVera:2022hyk}, together with 
  obtained or projected \textit{exclusion} limits from LHC/HL-LHC and LEPII (references
  can be found in \cite{ILCInternationalDevelopmentTeam:2022izu}).
  %%(cite ?).
  \textbf{NB:} At ILC, exclusion and discovery 
is almost the same, even for difficult channels.\label{fig:broadbrush}}
%\end{center}
\end{figure}

Although the LHC experiments have searched for and excluded many proposed new
particles up to masses in the 1 TeV range, there are many scenarios that are
difficult to address at a hadron collider \cite{Berggren:2020tle,Berggren:2015qua}.
%    \begin{itemize}
%    \item .
%    \item  {Complete coverage of Compressed spectra} - the most interesting case.
%    \end{itemize}
%\subsection* {Loop-hole free searches, even for low mass-differences}
In SUSY,  all is known for given masses, due to
  SUSY-principle, relating sparticle properties to those of the corresponding SM particles.
These relations do not depend on the SUSY breaking mechanism.
Obviously, there is one Next-to-lightest SUSY particle (NLSP), and it must have 100~\% BR
  to it's (on- or off-shell) SM-partner and the lightest SUSY particle (the LSP).
Therefore, one can perform  model independent evaluations of exclusion or discovery reach in $M_{NLSP} - M_{LSP}$ plane -
with no loop-holes.
%In addition,
One can do this for  any possible NLSP,
and concentrate on the "worst" cases - the ones with lowest cross-section and
most difficult signatures. These are on one hand the  bosinos, and on the other the  { $\stau$}.
The reaches for these cases are shown in Fig. \ref{fig:broadbrush}.
%\item  NLSP search $\leftrightarrow$ ``simplified models'' @ LHC!

%This talk will review a number of
%these scenarios and present the expectations for searches at an electron-positron
%collider such as the International Linear Collider. The cases discussed include
%the light Higgsino, the stau lepton in the coannihilation region relevant to
%dark matter, and heavy vector bosons coupling to the s-channel in e+e- annihilation.
%The studies are based on the ILD concept at the ILC.

%\begin{wrapfigure}{R}{8.5cm}

%%x \subsection{ILC SUSY measurements}
Since the discovery and exclusion reaches at the ILC are quite similar,
after a possible discovery, one will quickly enter the realm of precision
measurements.
%\subsection*{At ILC: discovery in a week...}
%A selectron signal after 5 fb$^{-1} \approx$ 1 week:
%  {\bf Bosinos}\\

\begin{figure}
%\begin{center}
\subfloat[][Natural SUSY]{
\includegraphics[scale=0.24]{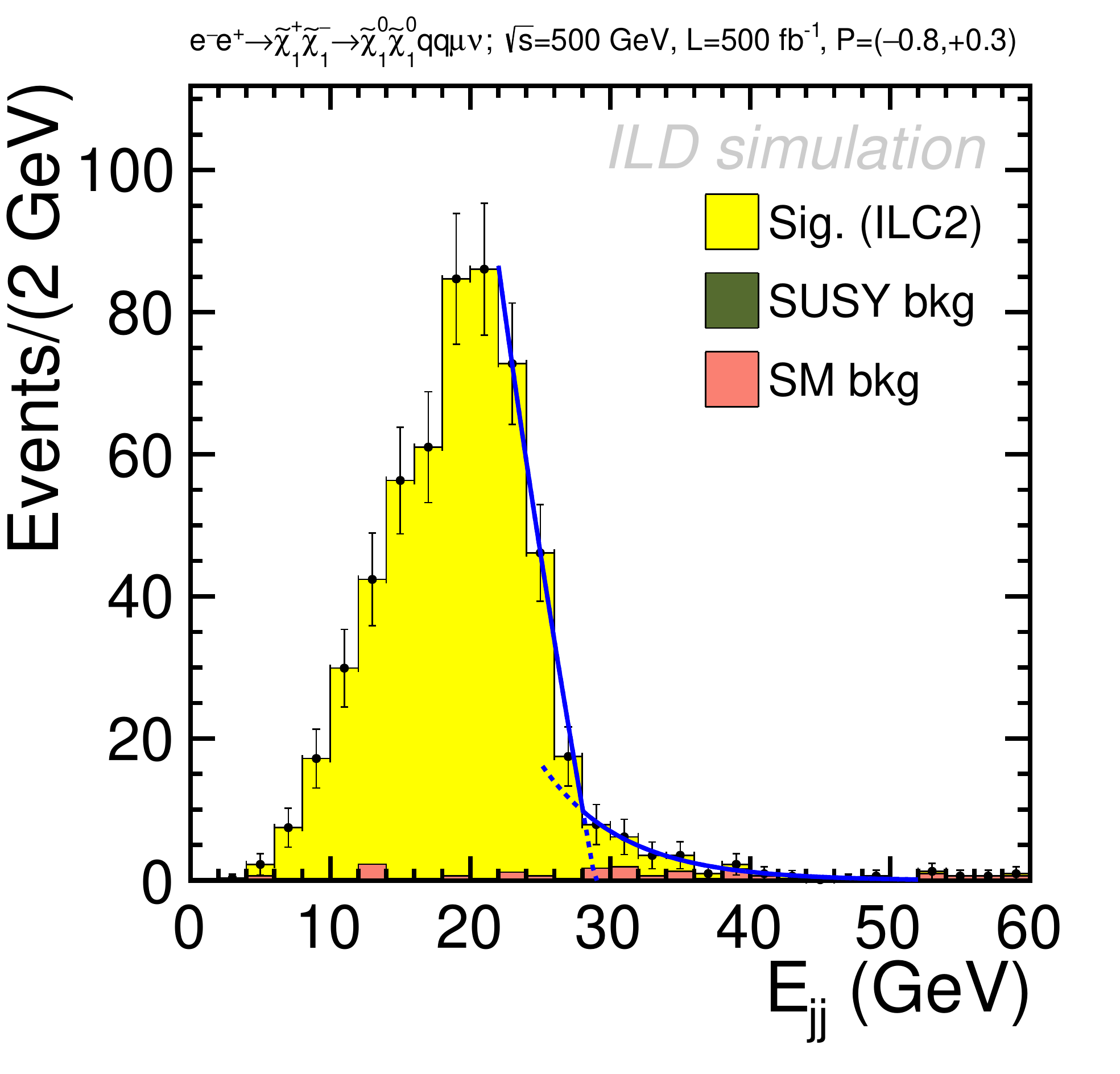}
}
\subfloat[][Extrapolation to GUT]{
\includegraphics[scale=0.24]{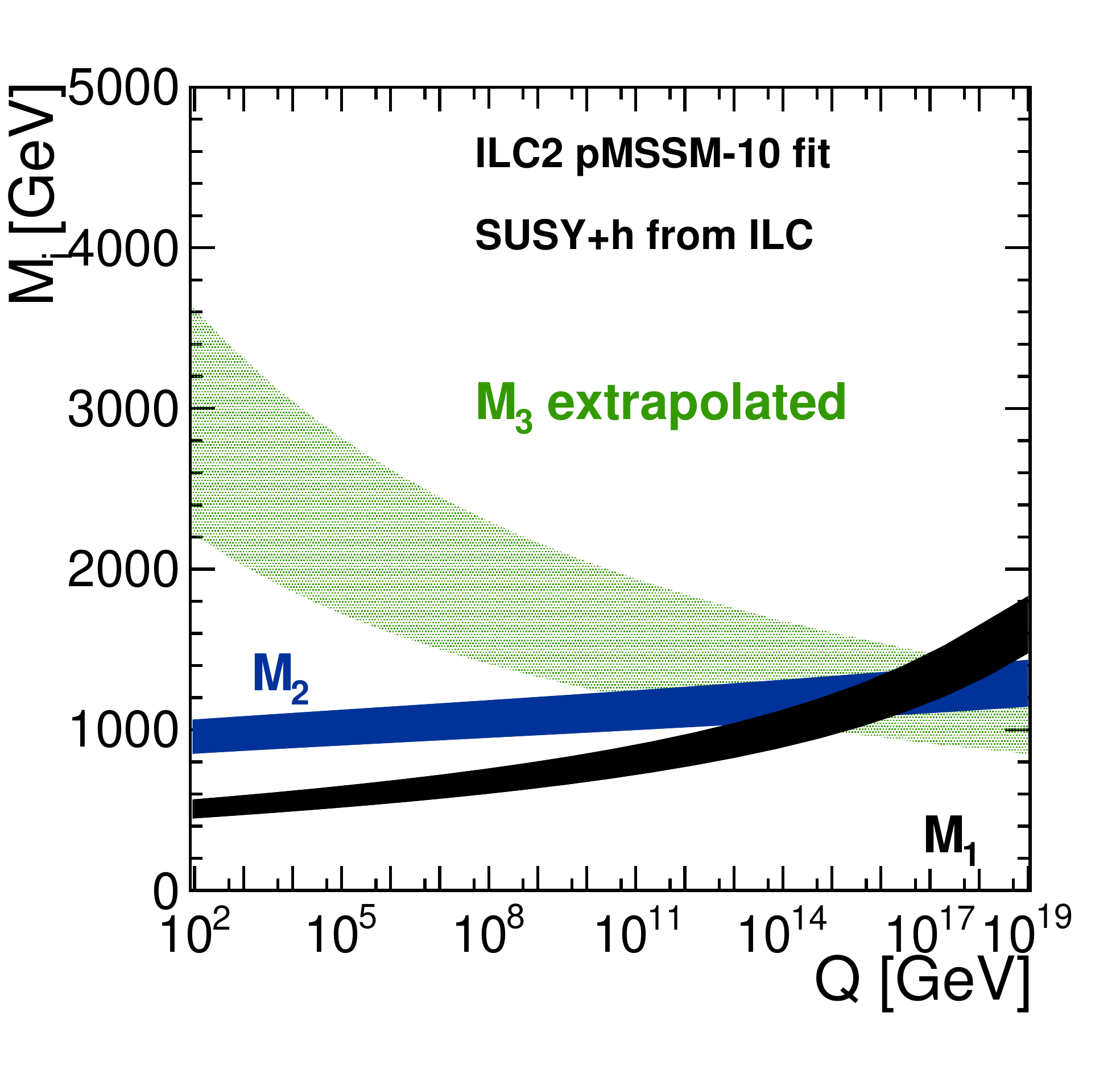}
}
\subfloat[][cosmology-motivated model]{
\includegraphics[scale=0.28]{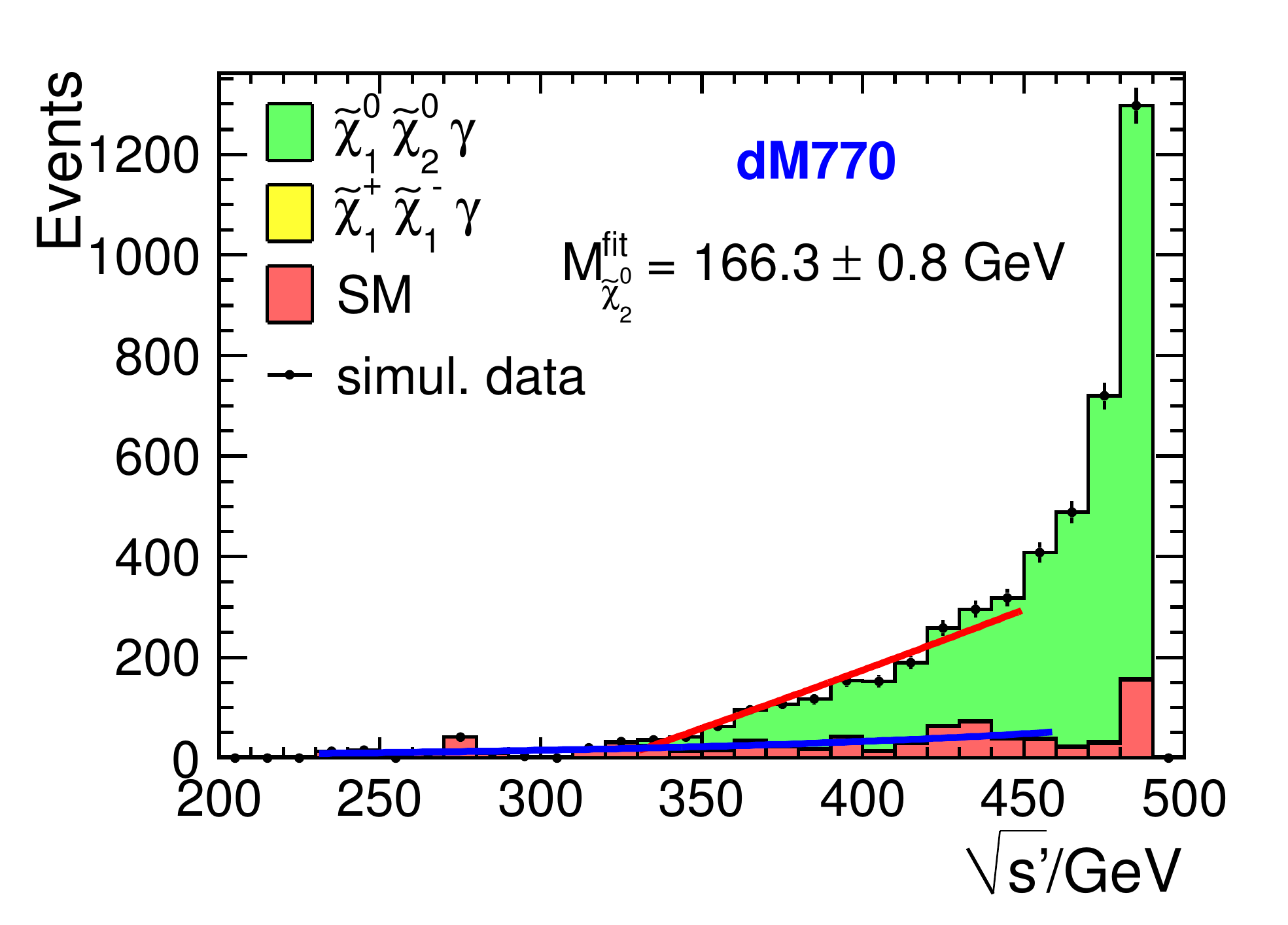}
}
\caption{ILC bosino measurements: (a) chargino signal in a natural SUSY model; (b)
RGE extrapolation of the gaugino mass-parameters to the GUT scale, with experimental inputs
obtained in natural SUSY; (c) neutralino signal in the cosmology-motivated model.
The SUSY background was included in the analysis, but
was found to completely removed after all cuts were applied.\label{fig:bosinos}}
%\end{center}
\end{figure}

In Fig. \ref{fig:bosinos} typical signals of bosino production are shown, for charginos (Fig. \ref{fig:bosinos}(a)) and 
neutralinos (Fig. \ref{fig:bosinos}(c)).
Both models are higgsino-LSP ones. The one to the left is one of
three studied natural SUSY models with  moderate mass
differences (15-20 GeV)  \cite{Baer:2019gvu}, while the one  to the right is a 
cosmology-motivated model, and has a sub-GeV difference \cite{Berggren:2013vfa}.
In the natural SUSY analysis, the combination of the measured masses, BR's and Higgs properties, all
10 weak-scale parameters gets constrained, for all three bench-marks.
In particular, the bino and wino SUSY breaking masses $M_1$ and $M_2$ - the ones most
directly related to the higgsino masses - can be determined at percent level.
The fitted weak-scale parameters can be evolved with the appropriate RGE's to higher scales.
This allows to verify or discard the idea of  GUT-scale unification of $M_1$ and $M_2$, see Fig \ref{fig:bosinos}(b).
%\begin{wrapfigure}{R}{8.5cm}
%typical By measuring 
%$\eeto \XP{1}+\XM{1} \rightarrow 
%(\ell\nu_{\ell}\XN{1})+(q\bar{q}^\prime\tilde{\chi}_1^0)$ 
%we are able to extract $\MXC{1}$
%and $\MXN{1}$ via the $m(jj)$ and $E(jj)$ distributions, 
%typically to percent level accuracy. Shown are

\begin{figure}[b]
%\begin{center}
\subfloat[][$\tilde{\tau}_1$]{
\includegraphics[scale=0.24]{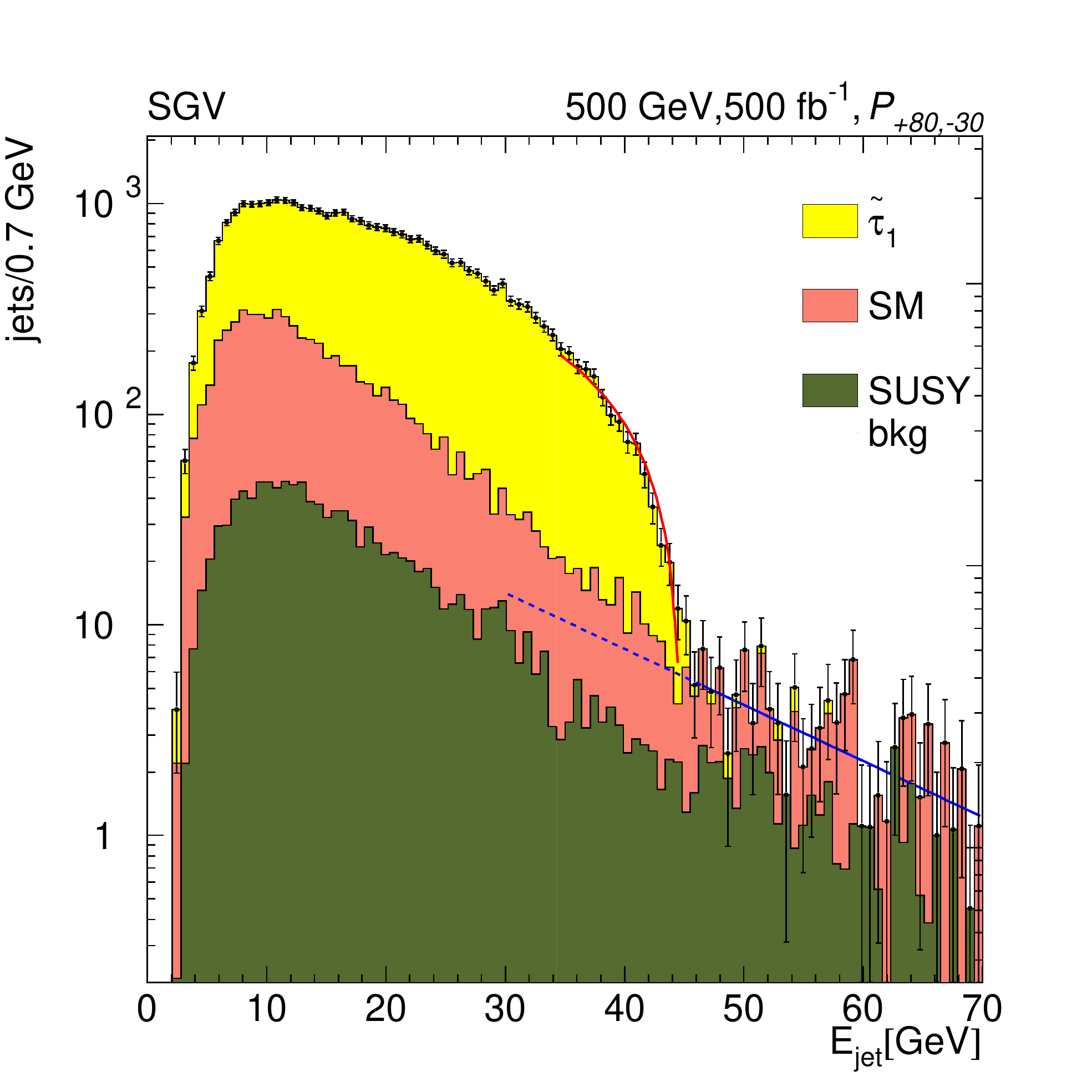}
}
\subfloat[][$\tilde{\mu}_R$]{
\includegraphics[scale=0.24]{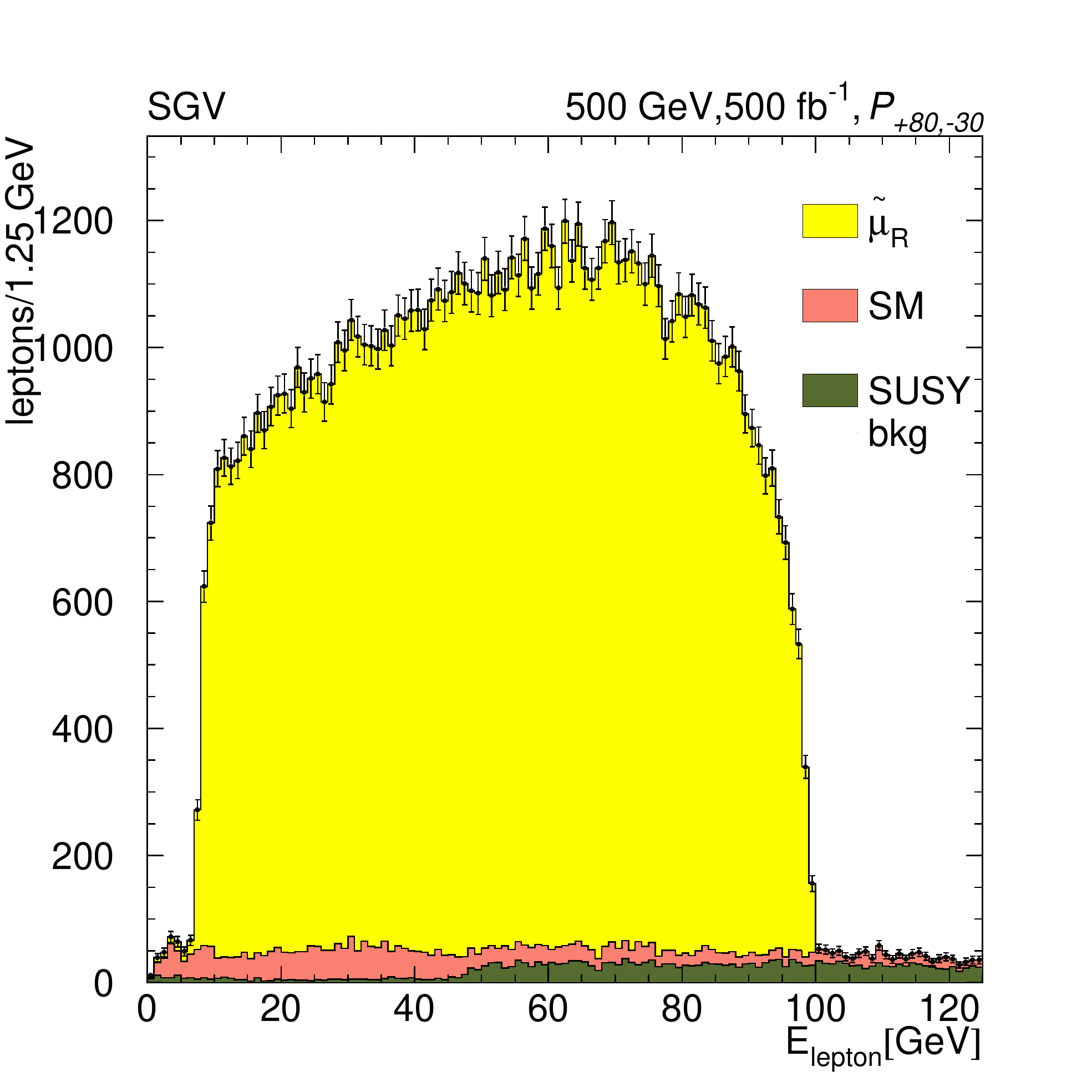}
}
\subfloat[][$\tilde{\mu}_R$ from cascades]{
\includegraphics[scale=0.24]{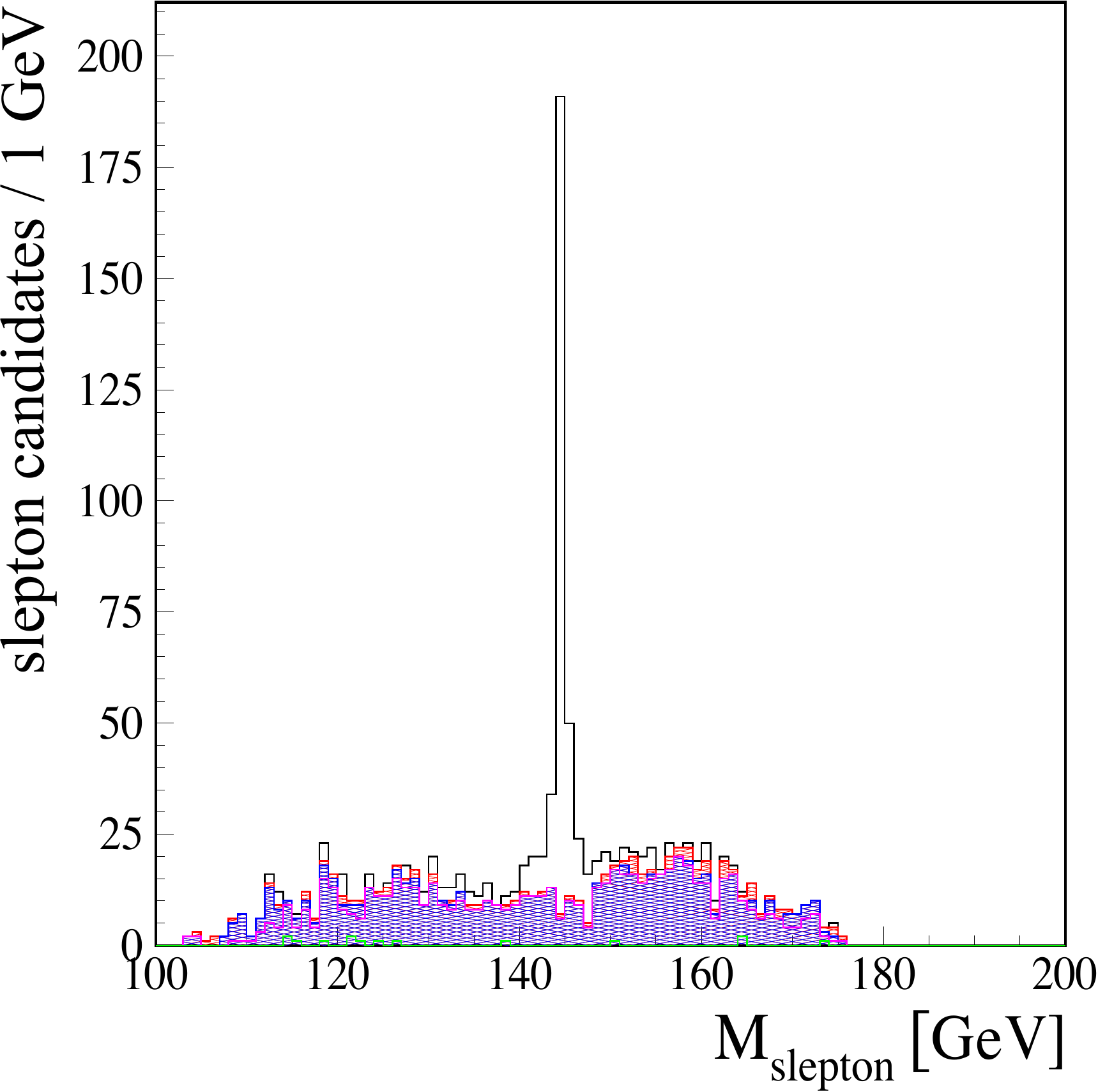}
}
\caption{Slepton signals at ILC: (a) $\tau$ decay-product spectrum in $\tilde{\tau}_1$ pair-production; 
(b) $\mu$ spectra in $\tilde{\mu}_R$ pair-production; (c) reconstructed $\tilde{\mu}_R$ mass
in a model where $\XN{2}$ decays to $\mu \tilde{\mu}_R$. Open: signal, red: SM background, blue: SUSY background. \label{fig:sleptons}}
%\end{center}
\end{figure}

In Fig. \ref{fig:sleptons}, typical slepton  signals are illustrated.
The end-points of the spectra of
the slepton decay-products,
enables to
measure the slepton and LSP masses to percent level accuracy.
Fig. \ref{fig:sleptons}(a) and (b) are examples of $\stone$ and $\smur$ signals, in a $\stau$ co-annihilation
model \cite{Berggren:2015qua}. Fig. \ref{fig:sleptons}(c) shows the signal that could be obtained from evaluating the
decay-kinematics in a model with cascade decays of $\tilde{\mu}$'s \cite{Berggren:2005gp}.
%\begin{wrapfigure}{R}{8.5cm}

After the full ILC program, and depending on model, channel, and polarisation, we find {experimentally} that
{measured $\delta (\mathrm{masses})$ = 0.5-1 \% , $\delta (\sigma \times \mathrm{BR})$ = 1-6\% }

\subsection{BSM at ILC: not only SUSY.}
\begin{figure}[t]
\begin{center}
\subfloat[][Mono-photon - heavy mediators]{
\includegraphics[scale=0.28]{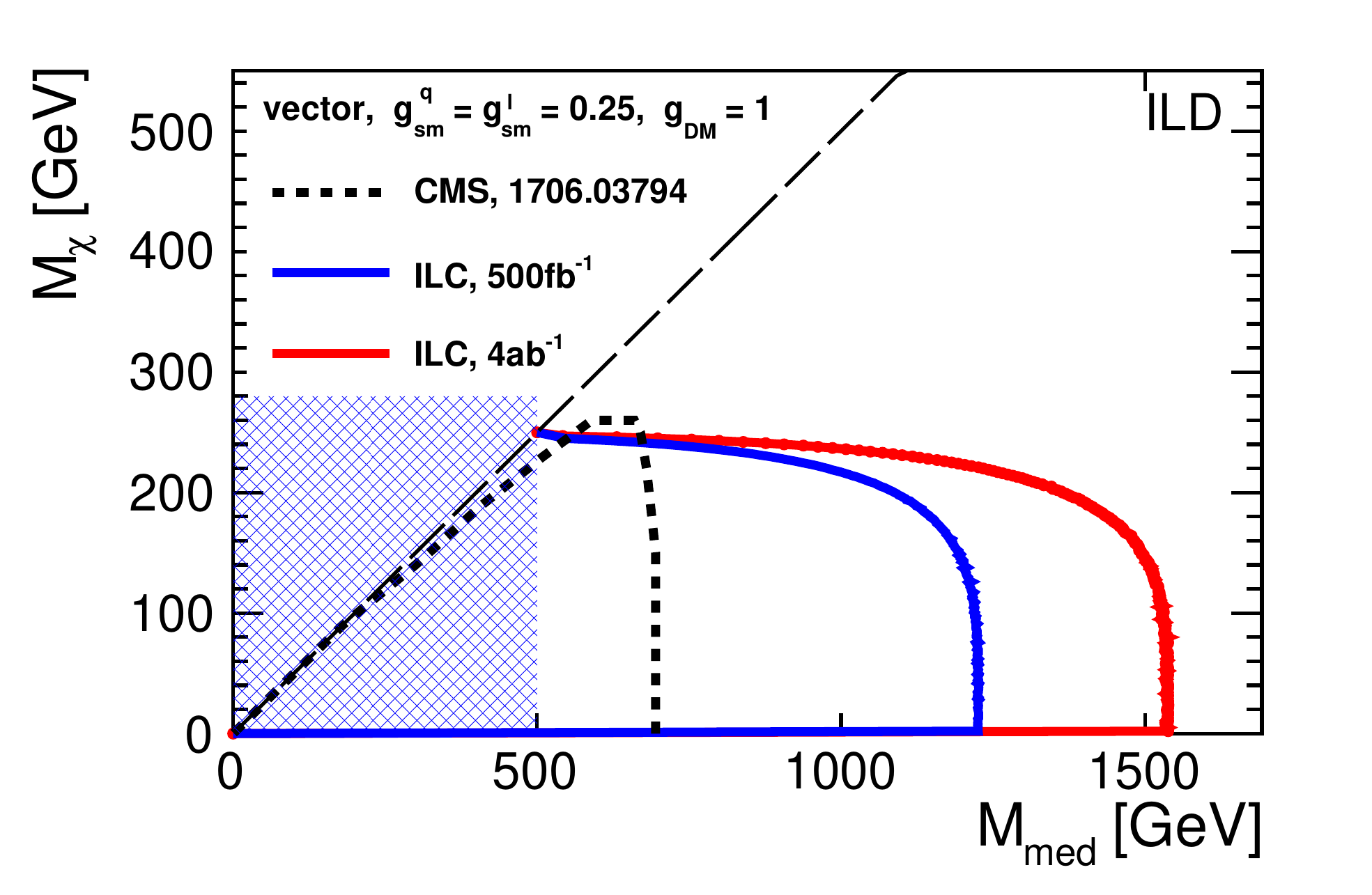}
}
%%x \subfloat[][Mono-photon - any mediator mass]{
%%x \includegraphics[scale=0.58]{plots/dm_light_med_ilc_fig11a}
%%x }
%%x 
\subfloat[][New scalar]{
\includegraphics[scale=0.24]{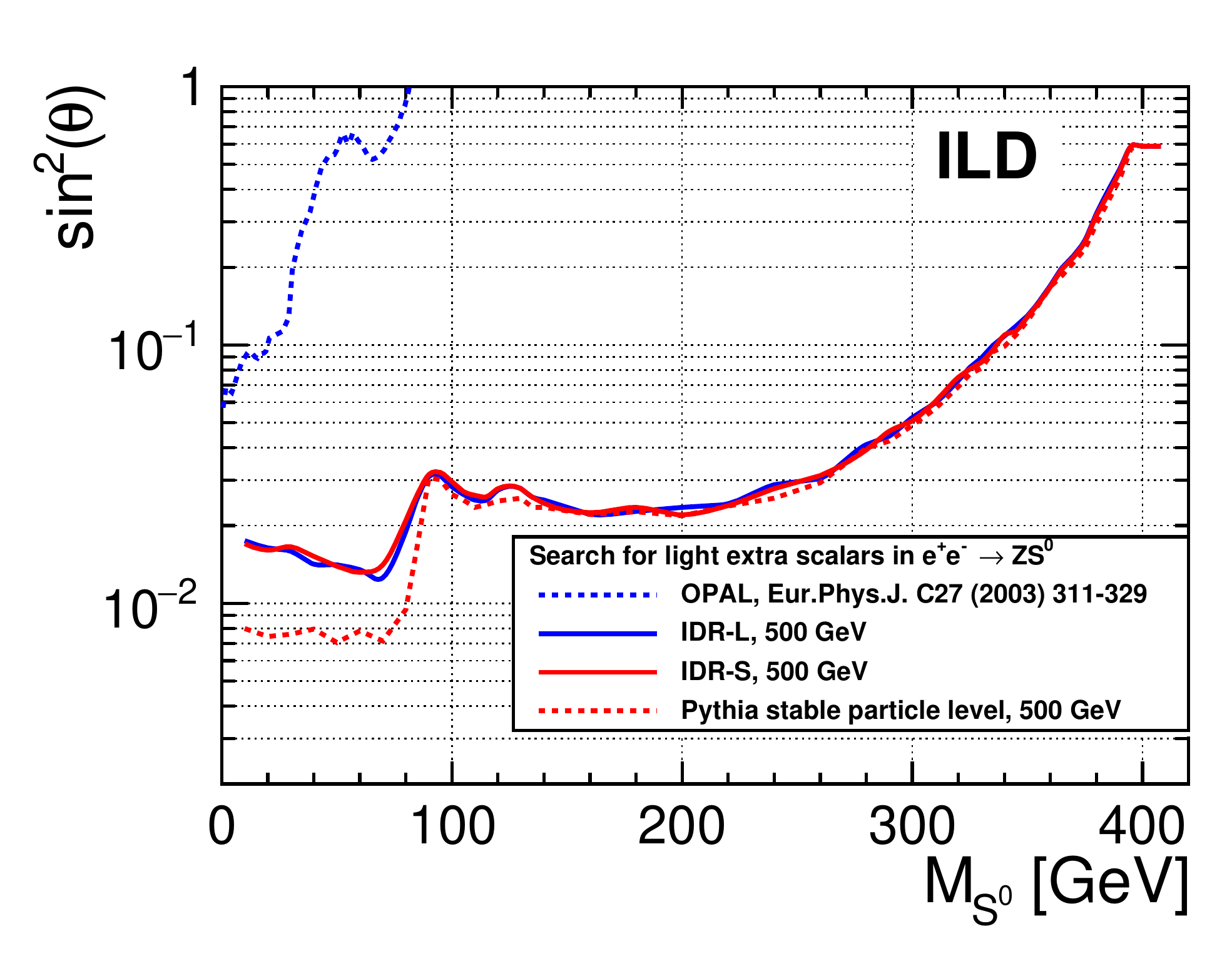}
}
\subfloat[][Dark photon]{
\includegraphics[scale=0.24]{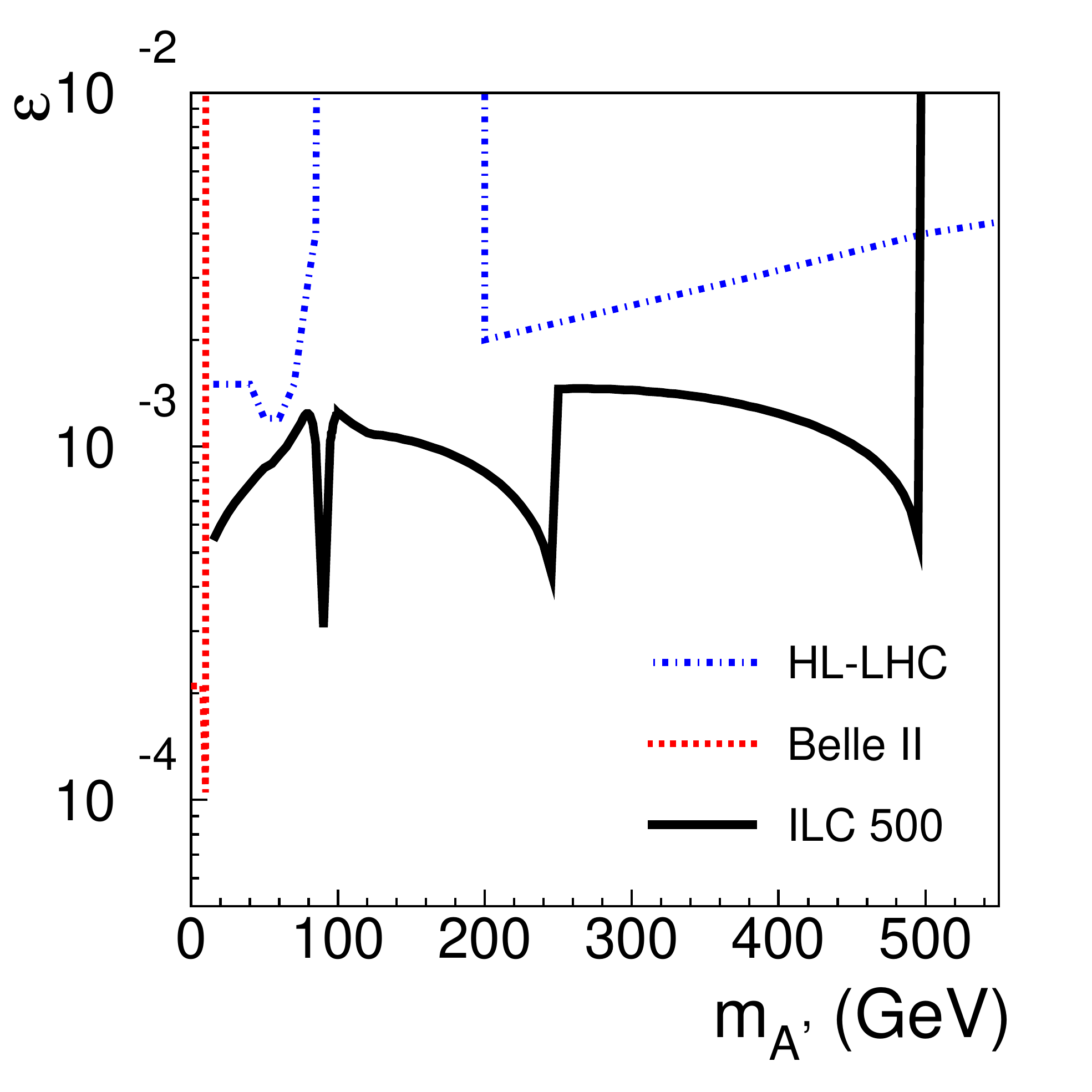}
}
%%x \subfloat[][Indirect BSM from SMEFT]{
%%x \includegraphics[scale=0.24]{plots/chi2plot_500}
%%x }
\caption{Examples of non-SUSY BSM searches at ILC\label{fig:otherbsm}}
\end{center}
\end{figure}

Dark matter can be searched for at ILC in $\eeto (DM)(DM) +\mathrm{ISR} \gamma$, i.e. in {Mono-photon} searches.
%%x Results of such searches are  shown in Fig. \ref{fig:otherbsm}(a) and (b), both for heavy mediators (a),
%%x where a model independet  EFT approach is appropriate, and for 
%%x arbitrary mediators (b), where the sensitivity will depend on properties of the
%%x mediator.
Results of such searches are  shown in Fig. \ref{fig:otherbsm}(a) for heavy mediators 
where a model independent  EFT approach is appropriate \cite{Habermehl:2020njb}. 
Also arbitrary mediators have been studied and also show potential beyond HL-LHC \cite{Kalinowski:2021tyr}
%\begin{wrapfigure}{R}{8.5cm}

New Higgs-like scalar (S),
  produced in $\eeto Z^* \rightarrow Z S$ with unknown decays of S, 
can be  searched for it in a decay-mode insensitive way at ILC, using the recoil-mass, i.e. 
the mass of the system recoiling against the measured $Z$ \cite{Wang:2020lkq}.
%    \item  {Example peaks} for a coupling equal to the an SM-Higgs at the same mass.
%               \tiny{(\href{https://arxiv.org/abs/2005.06265}{{\ttfamily arXiv:2005.06265}}})
    %\item
    Couplings down to a few percent of the  SM-Higgs equivalent can
    be excluded, as shown in Fig.  \ref{fig:otherbsm}(b).
%    \item  Note importance of {FullSim} !

At ILC one can also search for Dark photon/Z'. 
Generically, the kinetic mixing term $\frac{\epsilon}{2 \cos{\theta_W}} F^\prime_{\mu\nu}B^{\mu\nu}$ in the Lagrangian leads to 
 a tiny, narrow resonance, but still
      wide enough to make decays {prompt}. 
    One can search for this as a $\mu\mu$ resonance above background in $\eeto Z' + \mathrm{ISR} \rightarrow \mu^+ \mu^- + \mathrm{ISR}$. 
Results (from EPPSU) shown in Fig.  \ref{fig:otherbsm}(c).
%    \item Theory study, but with {reasonable assumption on resolution}.
%    FullSim study is W.I.P.
  %%  \item ILC compared to others (from EPPSU).

Another study that ILC allows for is indirect BSM.
Such studies do not only show an important discovery potential,
but also can achieve model separation.
%%x In Fig.  \ref{fig:otherbsm}(e), the results of an 
%%x SM effective field theory (SMEFT) study, using ILC results on Higgs properties and
%%x       TGCs are show.
%%x     Here, one did select models that are not discoverable at HL-LHC.
%%x The figure shows that, at ILC,  not only are the models separatable at 5 $\sigma$ from the SM,
%%x         but also from each other.
In \cite{Barklow:2017suo}  the results of an 
SM effective field theory (SMEFT) study, using ILC results on Higgs properties and
      TGCs are given
      %\tiny{(\href{https://arxiv.org/abs/1708.08912}{{\ttfamily Phys.~Rev.~D~97,0535003~(2018)}})}
    There, one selected models that are not discoverable at HL-LHC, and it was
shown that, at ILC,  not only are the models separatable at 5 $\sigma$ from the SM,
        but also from each other.
%%% %\begin{wrapfigure}{R}{8.5cm}
%%% \begin{figure}[b]
%%% %\begin{center}
%%% %\subfloat[][$\mu$ vs. $M_1$]{
%%% \includegraphics[scale=0.28]{plots/chi2plot_500}
%%% \caption{xxx\label{fig:smeft}}
%%% %\end{center}
%%% \end{figure}

\section{Conclusion}

 Sometimes, the capabilities for the direct discovery of new particles 
at the ILC exceed those
of the LHC, since ILC provides
a well-defined initial state,
a clean environment without QCD backgrounds,
%%% Democratic production of particles with electroweak charges. 
extendability in energy and polarised beams. 
Thanks to the low background levels, detectors - like ILD - can be factors more precise than their LHC counterparts,
can be hermetic, and do not need to be triggered.

Many ILC - LHC synergies from energy-reach vs. sensitivity are expected.
%\item {SUSY}: High mass vs. Low $\Delta(M)$. 
%\item 
E.g.  if {SUSY} is reachable at ILC, it means 5 $\sigma$ discovery, and precision measurements. This input
  might be just what is needed for LHC to transform a 3 $\sigma$ excess to a discovery of states
  beyond the reach of ILC.
Similar synergies can also be expected in many other models for BSM physics.
%  {both} ILC and LHC observes SUSY, the {(sub)percent} level measurements
%from ILC of the {lower states} will profit LHC to disentangle long decay-chains of
%{higher states}. 
%%x For models on Dark matter, on feebly interacting particles, or on new scalars,
%%x properties like whether the BSM physics is Leptophilic or Leptophobic, if it occurs at higher mass and features
%%x higher couplings compared to a case with lower mass and lower coupling also constitutes situations
%%x where ILC and LHC would be complementary.

%%xx \begin{itemize}
%%xx   \item Light higgsinos motivated by {naturalness}
%%xx   \item ILC would probe higgsinos {complementary to LHC reach}
%%xx   \begin{itemize}
%%xx     \item Either exclude masses up to $\sqrt{s}/2$=500 GeV for 1 TeV upgrade $\to$ wide coverage of natural SUSY scenarios
%%xx     \item or discover regardless of mass scale of heavier states 
%%xx   \end{itemize}
%%xx   \item ILC would measure properties of higgsinos to {sub-percent-level} precision.
%%xx %% with full ILC run / threshold scans to sub-\% precision
%%xx   \item Precise measurements allow for {\color{red}extracting GUT and weak scale} 
%%xx      parameters and {\color{red}predicting mass scales} of unobserved sparticles
%%xx \end{itemize}
%%xx \begin{center}
%%xx { \Huge
%%xx \color{desyorange}{ILC is {\it the} SUSY exploration instrument!}}
%%xx \end{center}

\section{Acknowledgements}
We would like to thank the LCC generator working group and the ILD software
working group for providing the simulation and reconstruction tools and
producing the Monte Carlo samples used in this study.
This work has benefited from computing services provided by the ILC Virtual
Organisation, supported by the national resource providers of the EGI
Federation and the Open Science GRID.

%%% biblatex
\printbibliography
\end{document}